\documentclass[letter,twocolumn]{jpsj3}
\usepackage{txfonts}
\usepackage{color}

\title{Evidence for 4D XY Quantum Criticality in $^4$He Confined in Nanoporous Media at Finite Temperatures}

\author{Tomoyuki Tani$^1$, Yusuke Nago$^1$, Satoshi Murakawa$^2$, and Keiya Shirahama$^1$}
\inst{$^1$Department of Physics, Keio University, Yokohama, Kanagawa 223-8522, Japan \\
$^2$Cryogenic Research Center, University of Tokyo, Bunkyou, Tokyo 113-0032, Japan}

\abst{ 
$^4$He confined in nanoporous media is a model Bose system that exhibits quantum phase transition (QPT) \textcolor{black}{by varying pressure.} 
We have precisely determined the critical exponent of the superfluid density of liquid ${}^{\rm 4}$He in porous Gelsil glasses with pore size of 3.0 nm using the Helmholtz resonator technique. 
The critical exponent $\zeta$ of the superfluid density was found to be $1.0 \pm 0.1$ for the pressure range $0.1 < P < 2.4$ MPa. 
This value provides decisive evidence that the finite-temperature superfluid transition belongs to the four-dimensional (4D) XY universality class, in contrast to the classical 3D XY one in bulk liquid $^4$He, in which $\zeta = 0.67$. 
The quantum critical behavior at a finite temperature is understood by strong phase fluctuations in local Bose-Einstein condensates above the superfluid transition temperature. 
$^4$He in nanoporous media is a unique example in which quantum criticality emerges not only at 0 K but at finite temperatures.
}

\begin{document}
\maketitle
Quantum phase transitions (QPTs), which are phase transitions between different ground states induced 
by quantum fluctuations, have been actively studied as a key phenomenon for understanding the physics of strongly correlated systems\cite{SondhiRMP1997,Sachdev2010}. 
In particular, QPTs have been discussed in connection with the interplay of magnetism and superconductivity in heavy-fermion systems and high-$T_{\mathrm c}$ cuprates\cite{GegenwartNPhys2008,KeimerNature2015}.
In such fermionic systems, however, it is difficult to achieve a full understanding of QPTs owing to their complexity and multiple degrees of freedom. 
In contrast, QPTs in bosonic systems are relatively simple and can be understood from a clear perspective based on the Bose-Hubbard model\cite{FisherPRB1989}, which is not only important in its own right but also useful for understanding fermion QPTs. 
However, there have been few real bosonic systems showing QPTs.

$^4$He in nanoporous media provides a unique example of a bosonic QPT\cite{YamamotoPRL2004,YamamotoPRL2008,ShirahamaJLTP2007,ShirahamaLTP2008,ShirahamaJPSJ2008}. 
The superfluid transition of bulk liquid $^4$He occurring at $T_{\lambda} \sim 2$ K is well-understood as a critical phenomenon in the 3D XY universality class. 
When liquid $^4$He is confined within a porous Gelsil glass with a pore size of a few nanometers, the superfluid transition temperature $T_{\mathrm c}$ decreases significantly\cite{YamamotoPRL2004}. 
As the pressure increases, superfluidity is further suppressed and $T_c$ reaches 0 K at a critical pressure $P_{\mathrm c}(0) \sim 3.3$ MPa. 
Thus, the confinement of $^4$He into nanospace produces a pressure-induced QPT characterized by a quantum critical point (QCP) at 3.3 MPa\cite{YamamotoPRL2004}.
The overall behavior of the QPT is qualitatively understood by the concept of localized Bose-Einstein condensates (LBEC)\cite{YamamotoPRL2008,ShirahamaJLTP2007,ShirahamaLTP2008,ShirahamaJPSJ2008}. 
Below the bulk $T_{\lambda}$, a number of nanoscale LBECs emerge in the pore voids, 
but they do not exhibit macroscopic superfluidity owing to the lack of phase coherence. 
Superfluidity occurs at a lower temperature at which phase coherence among LBECs is established. 

Eggel et al. discussed the QPT in terms of the Bose-Hubbard model, in which the LBECs are located at the 3D lattice sites\cite{EggelPRB2011}. 
They proposed that the QCP is governed by the 4D XY universality class, in which 4D consists of the spatial dimensions $d = 3$ plus a dynamical critical exponent $z = 1$. 
\textcolor{black}{The linear pressure dependence of the superfluid density at 0 K, $\rho _{\mathrm s}(P, T = 0) \propto P_{\mathrm c}(0) - P$, was explained as a quantum critical phonomenon.} 
Moreover, the theory \textcolor{black}{proposed the temperature dependence of the critical pressure at $T_{\mathrm c}$ as $P_{\mathrm c}(0) - P_{\mathrm c}(T) \propto T^{1/z\nu}$}, where $\nu$ is the critical exponent for superfluid coherence length.
The 4D XY value $\nu = 1/2$ gives $P_{\mathrm c}(0) - P_{\mathrm c}(T) \propto T^2$.
This was consistent with the experimental observation that the phase boundary is fitted to a power law $P_{\mathrm c}(0) - P_{\mathrm c}(T) \propto T^{2.13}$. 

In the general theory of QPT, however, a QCP is continuously connected to the finite-$T$ classical phase transition, which in the present case is governed by 3D XY criticality\cite{SondhiRMP1997}. 
Therefore, the $T^2$ dependence in the critical pressure suggests that quantum fluctuations dominate the superfluid transition even at finite $T$. 
To achieve a full understanding of the QPT, further experimental information on the finite-$T$ critical phenomenon is needed. 
In the present work, we have performed precise measurements of the critical exponent of superfluid density using the Helmholtz resonator technique\cite{AvenelPRL1985,Hoskinson2005,RojasPRB2015}. 
We discovered that the critical exponent of the superfluid density is unity over the pressure range that we could measure using this technique (0.1- 2.4 MPa). 
This result provides decisive evidence that the superfluid transition of $^4$He in Gelsil is governed by 4D XY quantum criticality even at finite temperatures. 
Our results establish $^4$He in nanoporous media as a unique bosonic model system in which strong quantum fluctuations dominate the system, not only at absolute zero but at finite temperatures as well.

\begin{figure}[t]
\centering
\includegraphics[width=1.0\linewidth]{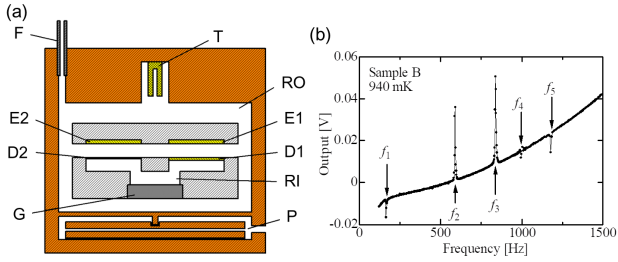}
\caption{\label{HR}(Color Online) (a) Schematic cross-sectional view of a Helmholtz resonator.
(D1) drive diaphragm, (D2) detector diaphragm, (E1) drive electrode, (E2) detector electrode, (G) porous Gelsil glass disk, (T) quartz tuning fork viscometer, (P) pressure gauge, and (F) liquid filling line. 
Liquid ${}^{\rm 4}$He is filled in the inner and outer volumes, and acts as helium reservoirs RI and RO. 
(b) Example of the spectrum of the Helmholtz resonator for Sample B. 
Data were taken at 940 mK. 
The five resonances are indicated by arrows. }
\end{figure}

Figure \ref{HR}(a) depicts the experimental apparatus.
The Helmholtz resonator consists of two superfluid reservoirs (RI and RO) separated by two circular diaphragms (D1 and D2) and a porous Gelsil disk (G). 
D1 and D2 are metal-deposited Kapton films whose thicknesses are 50 $\mu$m and 7.5 $\mu$m, respectively. 
Two fixed circular electrodes, located opposing D1 and D2, drives and detects the liquid flow through G.
A DC bias voltage of 350 V was applied to the diaphragms. 
A quartz tuning fork and a capacitive pressure gauge were installed for detecting superfluidity and measuring the pressure in bulk liquid ${}^{\rm 4}$He, respectively.
The measurements were performed at temperatures ranging from 0.7 to 2.5 K using a cryogen-free $^3$He refrigerator.

In this work, we have employed two Gelsil disks, which were cut out of rod samples of different batches. These are called Samples A and B. 
We measured the pore size distributions of the two disks applying the BJH method for N$_2$ adsorption - desorption isotherms\cite{BJH1951} and found that the 
two samples have identical pore size distributions. The distributions obtained from adsorption isotherms show a peak at 3.0 nm, while those obtained from desorption ones have a sharper peak at 3.8 nm, which indicates the size of bottlenecks in the porous structure. 
We adopted the former, 3.0 nm, as the representative pore diameter $d$ in this work. 
Samples A and B are both 9.0 mm in diameter and 1.0 and 2.0 mm thick, respectively. 
In the pore size distributions, Sample B has a slightly larger pore volume in the size range $1.5 < d < 3.0$ nm compared to Sample A. 

We measured the flow characteristics as follows: 
The liquid RI is mechanically driven by oscillating the diaphragm D1 using an AC voltage applied to the electrode E1.
The state of flow in and between RI and RO is detected from the oscillations of D2, which is much thinner than D1, by monitoring the AC voltage induced in the electrode E2 using a current preamplifier and lock-in amplifier. 
We first performed a measurement of the spectrum as a function of the driving frequency. 
An example of the spectrum is shown in Fig.~\ref{HR}(b). 
Five resonant modes are observed at frequencies ranging from 10 to 1500 Hz, and there are several resonances above 2000 Hz.
In this work, we have examined $T$ dependencies of the resonant frequencies and the linewidth (dissipation) of the two lowest modes.

When the system is above the superfluid transition temperature of $^4$He in Gelsil, $T_{\rm c}$, liquid ${}^{\rm 4}$He is blocked in Gelsil owing to its viscosity. 
When $T_{\rm c} < T < T_{\lambda}$, each resonance is dominated by superfluidity and compressibility of the liquid inside RI. 
Below $T_{\rm c}$, liquid flows through Gelsil as a superfluid. 
The superflow changes the resonant frequencies $f_n$ and amplitudes by the reduction of mass loading and changes in the dissipation of liquid $^4$He.
Here, the index $n = 1, 2, 3, ...$ denotes the order of the modes counted from lowest frequency mode. 


\begin{figure}[tb]
\centering
\includegraphics[width=1.0\linewidth]{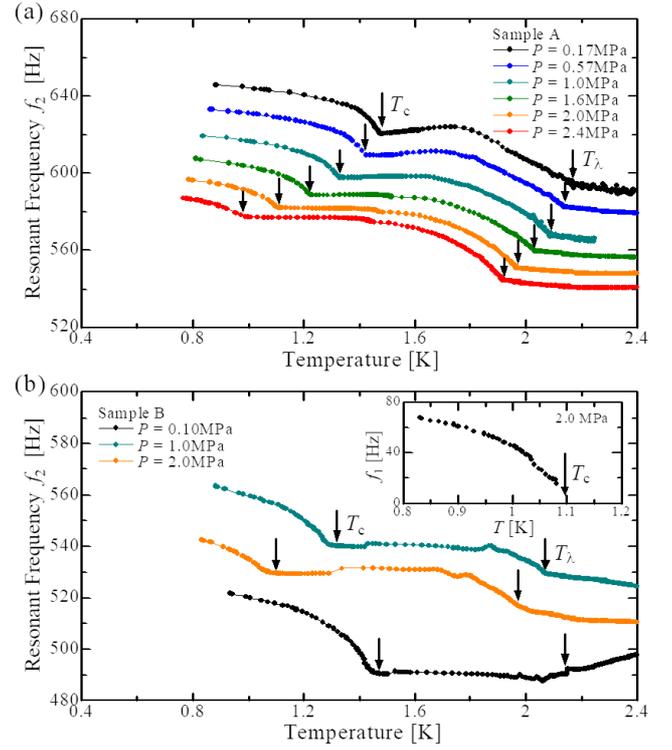}
\caption{\label{freq}(Color Online) Resonant frequencies of the second mode $f_2(T)$ at different pressures. 
(a): Sample A, (b): Sample B. 
$T_{\rm\lambda}$ and $T_{\rm c}$ indicate superfluid transition temperatures in the bulk and in the Gelsil, respectively. 
Inset \textcolor{black}{in (b)} shows the resonant frequency of the first mode $f_1(T)$ at $P = 2.0$ MPa. 
}
\end{figure}

In Fig.~\ref{freq}, we present the temperature dependence of the resonant frequency $f_{2}(T)$ at various pressures. 
As $T$ decreases, $f_{2}$ increases at $T_{\rm\lambda}$ and $T_{\rm c}$. 
The increase in $f_{2}$ below $T_{\lambda}$ is attributed to the reduction of viscous friction in the RI. 
At $T_{\rm c}$, the liquid inside Gelsil nanopores undergoes superfluid transition, resulting  
in a sharp increase in $f_{2}$ by the sudden reduction in the effective liquid mass participating in oscillation. 
The mass reduction should be proportional to the superfluid density of $^4$He in Gelsil. 

We analyze the resonance at $f_2$ as follows:
The resonance characteristics indicate that the two diaphragms D1 and D2 oscillate \textit{in phase}.
This is because liquid $^4$He in RI can be practically regarded as incompressible, i.e. its compressibility is negligibly small. 
Therefore, the liquid in RI participates coincidently \textcolor{black}{with} the oscillations of two diaphragms.
Then, a simplified model for the resonant mode is applied by considering the oscillation of the liquid \textcolor{black}{under the restoring forces from the two diaphragms having parallel spring constants. }
Assuming that the velocity of $^4$He in Gelsil is less than the superfluid critical velocity, the resonant frequency $f_2$ below $T_{\rm c}$ is given by
\begin{equation}
  f_{\rm 2} = \frac{\rm 1}{\rm 2\it\pi}\sqrt{\frac{\rm 8\it\pi(\sigma_{\rm 1} + \sigma_{\rm 2})}{\rho V - \alpha\rho_{\rm s}}}. 
  \label{reso}
\end{equation}
Here $\sigma_{\rm 1}$ and $\sigma_{\rm 2}$ are the tensions of the diaphragms D1 and D2, respectively, $\rho$ is the density of the bulk liquid ${}^{\rm 4}$He, $\rho_{\rm s}$ is the superfluid density of $^4$He in Gelsil, $V$ is the total volume of the inner and outer liquid, and $\alpha$ is a coefficient that depends on the configuration of Gelsil.

As shown in Fig.~\ref{freq}, $f_{2}$ decreases across the entire $T$ range as $P$ increases owing to the increase in liquid mass. 
An exception is shown in Fig.~\ref{freq} (b), in which $f_{2}$ at 0.10 MPa is lower than those at higher pressures. We attribute this behavior to an unstable characteristic in the tension of D2. 
Despite such an instability, the behavior below $T_{\rm c}$ is stable. 

The superfluid density, $\rho_{\rm s}$, of $^4$He in the Gelsil is determined from $f_2$ by
\begin{equation}
\frac{\rm 1}{f_{\rm 2}^{\rm 2}(T_{\rm c},\rm 0)} - \frac{\rm 1}{f_{\rm 2}^{\rm 2}(T,\rho_{\rm s})} = \frac{\pi\alpha}{\rm 2(\it\sigma_{\rm 1} + \sigma_{\rm 2})}\rho_{\rm s}.
\label{sd}
\end{equation}
Here, the density of liquid ${}^{\rm 4}$He is regarded as constant in the current temperature range. 
However, the measured resonant frequencies just above $T_{\rm c}$ have a small but finite $T$ dependence, as shown in Fig.~\ref{freq}.
We subtract this as a background offset assuming a linear $T$ dependence. 
Assuming that the tensions $\sigma_{\rm 1}$ and $\sigma_{\rm 2}$, and the structural coefficient $\alpha$ are constant, the left-hand side of Eq.~(\ref{sd}) determines the $T$ dependence of 
$\rho_{\rm s}$. 
A similar analysis can be applied to the third resonance $f_3$.

We found that the lowest frequency resonance $f_1$ occurs by a different mechanism. 
A data set of $f_1(T)$ is shown in the inset of Fig.~\ref{freq}(b). 
The lowest mode is observed only at $T < T_{\rm c}$, as the dissipation rapidly increases in the vicinity of $T_{\rm c}$. 
This behavior is in contrast with the other higher modes at $f_2$ and $f_3$, in which the resonances are identified up to 2.5 K.
We consider the mechanism of the $f_1$ mode as follows:
When the drive diaphragm D1 pushes the superfluid inside the RI, some of the liquid flows from RI to RO via Gelsil.
This flow pushes down the detection diaphragm D2 which is much more flexible than D1.
This mechanism is confirmed by the fact that D1 and D2 oscillate \textit{out of phase} at $f_1$, as opposed to the higher modes $f_2$ and $f_3$. 
We then obtain $f_{1}(T)$, by applying a derivation in the case that the resonator has a single diaphragm\cite{Hoskinson2005,RojasPRB2015}, as a formula  
\begin{equation}
  f_{1}(T) = \frac{\rm 1}{\rm 2\it\pi}\sqrt{\frac{\rho_{\rm s}(T)}{\rho^{\rm 2}} \beta \frac{\rm 8\it\pi({\rm 1}/\sigma_{\rm 1}+{\rm 1}/\sigma_{\rm 2})^{\rm -1}}{A_{\rm d}^{\rm 2}}}.
\label{lfm}
\end{equation}
Here, $\beta$ is a coefficient that depends on the configuration of the channel, and the liquid $^4$He is regarded as imcompressible. 
We find that $\rho_{\rm s}(T)$ calculated from Eq.~(\ref{sd}) agrees well with Eq.~(\ref{lfm}).
This provides further support for the validity of the determination of $\rho_{\mathrm s}$ using Eq.~(\ref{sd}). 
The details of the analyses will be presented elsewhere\cite{Tani2021}.

In order to discuss the critical phenomenon, we analyze $\rho_{\rm s}$ in the vicinity of the superfluid transition using a power law:
\begin{equation}
  \rho_{\rm s} \propto \left|\rm 1 -\it T/T_{\rm c}\right|^{\zeta},
\end{equation}
where $\zeta$ is the critical exponent for superfluid density. 

\begin{figure}[tb]
\includegraphics[width=1.0\linewidth]{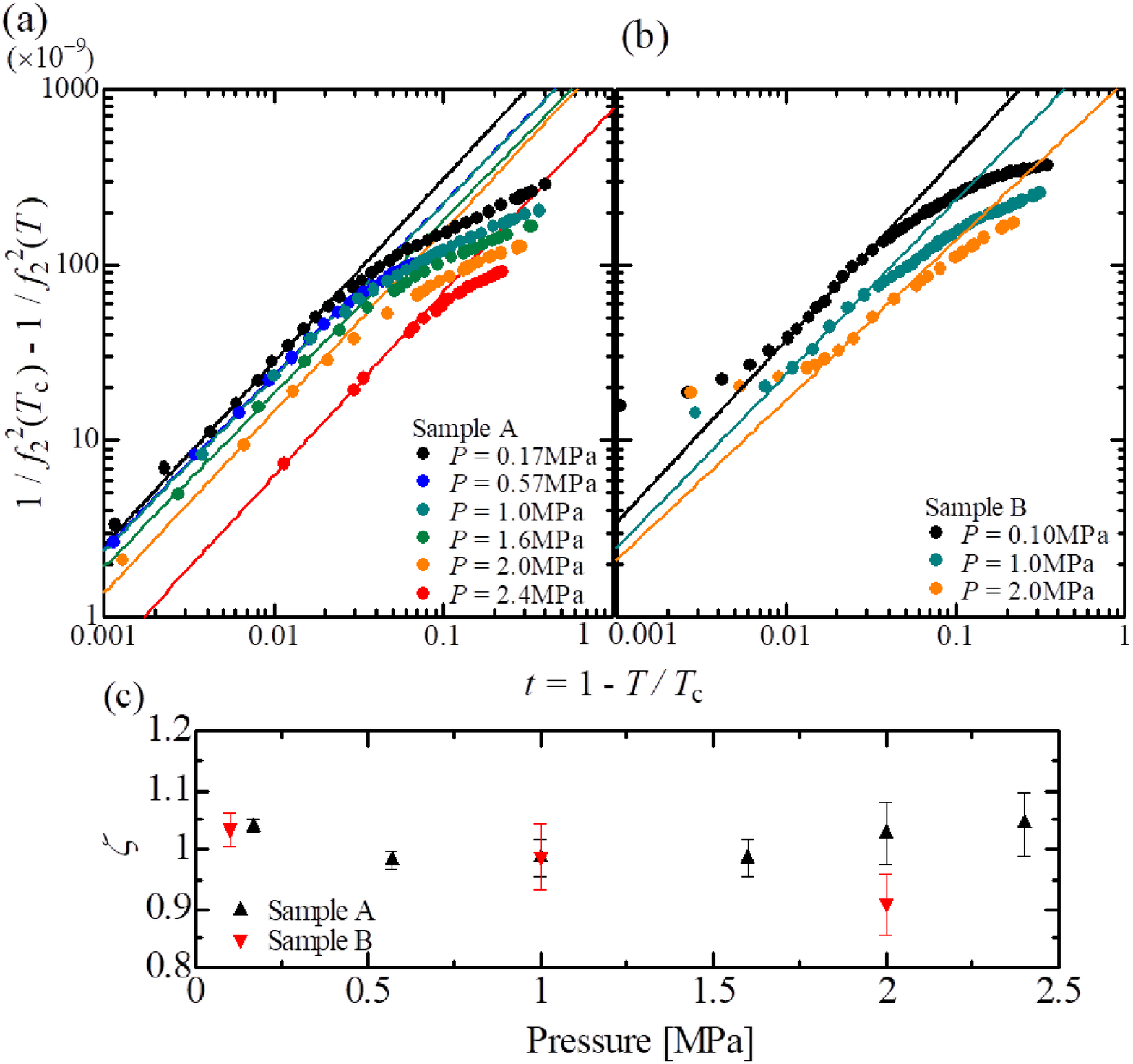}
\caption{\label{ce}(Color Online) Log-log plots of $1/{f_2}^2(T_{\mathrm c}) - 1/{f_2}^2(T)$, which is proportional to the superfluid density $\rho_{\rm s}$, as a function of the reduced temperature $t = 1 - T/T_{\mathrm c}$ under different pressures. \textcolor{black}{(a): Sample A, (b): Sample B.} 
(c) Critical exponents $\zeta$ versus pressure for each Gelsil sample.}
\end{figure}

In Fig.~\ref{ce}, we present $\log$-$\log$ plots of the quantity $1/{f_2}^2(T_{\mathrm c}) - 1/{f_2}^2(T)$, which is proportional to $\rho_{\rm s}$, obtained from the data shown in Fig.~\ref{freq}, as a function of the reduced temperature $t = 1 - T / T_{\rm c} $. 
In these plots, we determined $T_{\rm c}$ to make the longest straight lines in the log-log plots. 
We see in Fig.~\ref{ce}(a) that the data \textcolor{black}{of Sample A} are well fitted by straight lines in the range  $0.001 < t < 0.03$, while 
the plots for \textcolor{black}{Sample B} shown in Fig.~\ref{ce}(b) deviate upward from straight lines at $t < 0.01$. 
We discuss the possible origin of the deviation later. 
The critical exponent $\zeta$ is obtained from the slope of the straight lines. 
Figure \ref{ce}(c) shows $\zeta$ as a function of the pressure.
We found that $0.91 < \zeta < 1.04$ at all pressures and in the two different Gelsil samples. 
No pressure dependence was observed. 

The critical exponents of $^4$He in Gelsil not only differ from the bulk superfluid critical exponent $\zeta_{\mathrm b} = 0.67 (= 2/3)$\cite{AhlersRMP1980} but are also larger than $\zeta$ of $^4$He confined in other porous materials: 
In porous Vycor glass, $\zeta \sim 0.67$\cite{KiewietHallReppy}, which is the same as the bulk one, whereas in $^4$He in aerogel and in xerogel\cite{ChanPRL1988}, in which the pore sizes  ($> 100$ nm) and porosity ($90 \sim 98$ \%) are much larger than those of our Gelsil samples, $0.8 < \zeta < 0.9$, depending on the material and sample batch.
The superfluid transition temperatures in these porous materials are located between 1.95 and 2.10 K (at SVP), very close to the bulk $T_{\lambda}$, and do not exhibit any QPT at high pressures. 
Therefore, the superfluid transitions 
in these porous materials are classical, unlike $^4$He in Gelsil. 

\begin{figure}[tb]
\centering
\includegraphics[width=1.0\linewidth]{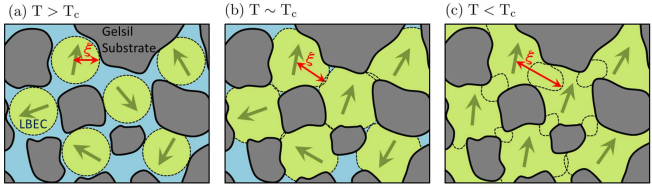}
\caption{\label{LBECIllustration}(Color Online) Illustration of the mechanism of superfluid transition. Phases of the order parameter of LBECs are depicted symbolically by arrows. (a) When the coherence length $\xi$ is shorter than the mean pore length $l$, which is roughly comparable to the pore diameter $d$, the phases fluctuate and have no coherence between LBECs. (b) When $\xi$ grows and reaches $l$, the phases begin to correlate among LBECs. This correlation determines $T_{\mathrm c}$. (c) Well below $T_{\mathrm c}$, the BECs govern the entire system, and phase coherence is established. }
\end{figure}

We argue that $\zeta = 1$ in $^4$He in Gelsil is attributed to a strong quantum fluctuation. 
The mechanism of superfluid transition is illustrated in Fig. \ref{LBECIllustration}.
At temperatures $T_{\mathrm c} < T < T_{\lambda}$, each LBEC whose size is comparable with the superfluid coherence length $\xi$ has different phases (Fig. \ref{LBECIllustration}(a)). 
As $T$ decreases, $\xi$ increases, and the whole system undergoes superfluid transition when $\xi$ reaches the mean length of nanopores $l$, which is comparable to the distance between LBECs (Fig. \ref{LBECIllustration}(b)). 
The pore length $l$ is comparable to or slightly larger than the mean pore diameter $d$, depending on the porosity of the material. 
In the superfluid state well below $T_{\mathrm c}$, $\xi$ exceeds $l$, and phase coherence is established, as illustrated in Fig. \ref{LBECIllustration}(c).

The phase coherence between LBECs is \textcolor{black}{realized} by frequent exchanges of $^4$He atoms between them. 
In $^4$He in Gelsil, this particle exchange is strongly suppressed by the narrowness of the pores. 
When a $^4$He atom is transferred from an LBEC at the $i$-th site to its nearest neighbor, the increase in energy (i.e. the charging energy analogous to the superconducting Josephson junction) is given by 
$V_{i} = (\mathcal{V}_{i}\nu^{\rm 2}\kappa)^{-1}$, where $\mathcal{V}_{i}$ is the volume of the \textcolor{black}{$i$-th} site, $\nu$ is the number density, and $\kappa$ is the compressibility of $^4$He\cite{EggelPRB2011}. 
\textcolor{black}{When $T_{\mathrm c} < V_{i}$, the transition is governed by quantum fluctuations and hence the system belongs to the $d + z = 4$ dimensional universality class.} 
Using \textcolor{black}{the bulk $^4$He compressibility for} $\kappa$, and assuming that the two atomic layers adjacent to the pore wall are localized and do not participate in superfluidity\cite{Makiuchi2018}, $V_{i}$ is estimated to be 0.54 K. 
This value is smaller than the previous estimation, 1.4 K\cite{EggelPRB2011}, but in reality $V_{i}$ may be several times 0.54 K, as it is suggested that the compressibility of \textcolor{black}{a liquid is suppressed by confinement into nanopores}\cite{GorJCP2015}. 
It is therefore suggested that all the superfluid transitions observed in the present work are dominated by quantum fluctuations, and the system obeys 4D XY criticality. 

In the $\log - \log$ plot for Sample B (Fig.~\ref{ce}(b)), deviations from the power law were observed at $t < 0.01$. 
This reflects that  the $f_{2}(T)$ data of Sample B are rounded at $T_{\mathrm c}$ compared to those of Sample A, as shown in Fig.~\ref{freq}(a) and (b).  
Such rounding was also observed in $^4$He confined in other porous media such as Vycor and Aerogel\cite{ChanPRL1988}. 
The rounding (tail) in $\rho_{\mathrm s}$ might imply a distribution of $T_{\mathrm c}$ in a single sample. 
In the present work, two Gelsil samples have identical pore size distributions, except that Sample B has a slightly larger pore size distribution at $d < 3$ nm compared to Sample A. 
However, as smaller pores could further lower $T_{\mathrm c}$, this difference is not likely to produce the distribution of $T_{\mathrm c}$ at higher temperatures. 
We speculate that the thickness of glass samples is related to the rounding in Sample B. 
Superflow via Gelsil occurs only when the LBEC phases become coherent throughout the Gelsil sample. 
At a temperature very close to $T_{\mathrm c}$, phase coherence may occur probabilistically on a macroscopic scale. 
Longer samples may therefore have a lower probability of realizing a firm phase coherence from one reservoir to another.   
Length-scale dependence on a QPT has been observed in 1D Josephson junction arrays, in which the system shows a 2D XY universality class\cite{Haviland1998}. 
Helmholtz measurements using thinner Gelsil samples will clarify the origin of the sample dependence.

The superfluid order parameter $\Psi$ near $T_{\mathrm c}$ is expressed by $|\Psi| \propto (1-T/T_{\rm c})^{\beta}$. 
The mean field theory, which is valid in 4D XY, gives the critical exponent $\beta = 0.5$.
As the superfluid density $\rho_{\rm s}=|\Psi|^{2}$, the critical exponent for $\rho_{\rm s}$ is $\zeta = 2\beta = 1$.
This agrees with the critical exponents shown in Fig. \ref{ce}(c).

The $T$ dependence of $\rho_{\rm s}$ was also examined in a previous torsional oscillator (TO) study\cite{YamamotoPRL2004}. 
However, analyses of critical exponents were disturbed by the contribution of the bulk superfluid component to the resonant frequency data at $P < 2.5$ MPa, where bulk $^4$He is liquid, and by the coupling of torsional oscillation to a superfluid sound resonance at $P > 2.5$ MPa. 
The present work revealed for the first time the 4D XY critical behavior in superfluid density at finite temperatures.

\begin{figure}[tb]
\centering
\includegraphics[width=0.8\linewidth]{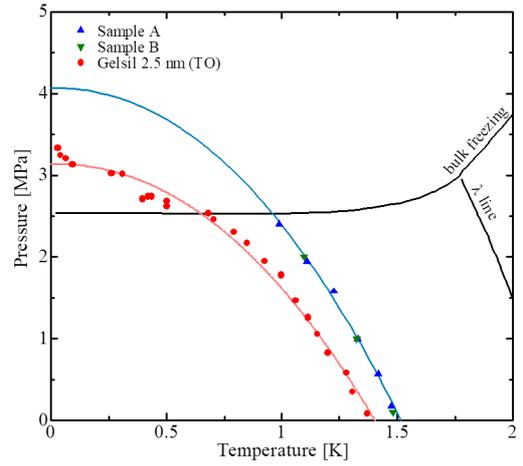}
\caption{\label{pd}(Color Online) Superfluid phase boundary. The triangles and inverted triangles denote Sample A and B, respectively. 
Circles indicate the $T_{\rm c}$ data determined in the previous torsional oscillator (TO) study for a Gelsil sample of a different batch\cite{YamamotoPRL2004}. 
The results of fittings to power law $P_{\mathrm c}(0) - P_{\mathrm c}(T) \propto T^{2.13}$ are indicated by solid lines.
Other solid lines show the bulk freezing curve and the superfluid $\lambda$ line. }
\end{figure}

Figure \ref{pd} shows the $P$-$T$ phase diagram 
together with the previous result\cite{YamamotoPRL2004,YamamotoPRL2008}.  
$T_{\rm c}$ is 1.45 K at 0.1 MPa, and decreases to 0.98 K at 2.4 MPa. 
In this work the phase boundary is obtained up to 2.4 MPa, because above 2.5 MPa bulk $^4$He freezes in the resonator.  
$T_{\rm c}$ is located at temperatures higher than those for the previous result.
Although this difference is not yet understood, similar differences have also been observed in an experiment using a Gelsil sample from the same batch as the TO study\cite{KobayashiJPSJ2010}. 
We therefore speculate that some unknown difference in the porous structures \textcolor{black}{such as the size of the pore bottlenecks} is essential for the location and curvature of the $T_{\mathrm c}$ lines. 
Assuming the result of \textcolor{black}{the powerlaw fitting 
by} Eggel et al., $P_{\mathrm c}(0) - P_{\mathrm c}(T) \propto T^{2.13}$, the $T_{\mathrm c}$ lines can be fitted with $P_{\mathrm c}(0) = 4.08$ MPa. 
This value is higher than \textcolor{black}{the freezing pressure 
determined} in a previous measurement, $P_{\mathrm f} \sim 3.7$ MPa\cite{YamamotoJPSJ2007}. 
In Samples A and B, the QCP may be masked by the solid phase. 
Such masking, even if it exists, does not influence the quantum critical nature seen at lower pressures.

Finally, we comment on the \textcolor{black}{dissipation 
obtained} from the resonance linewidths. 
An excess dissipation peak was observed just below $T_{\mathrm c}$.
We performed a Helmholtz measurement using a Vycor glass, which had a pore diameter of 6.6 nm. 
The superfluid transition was observed at 1.97 K (at $P = 0.1$ MPa) as a sharp increase in the resonance frequency, but dissipated abruptly \textit{decreased} just below $T_{\mathrm c}$, in contrast to the \textit{increase} in Gelsil. 
We attribute the excess dissipation seen in Gelsil to the energy produced by the orientation of the phases of the LBEC order parameter. 
When many LBECs are connected at $T_{\mathrm c}$, mismatches in phases must be eliminated keeping the superfluid circulations quantized in any closed path in liquid $^4$He. 
Below $T_{\mathrm c}$, the phases of the order parameter match continuously in helium in the pores and in the two superfluid reservoirs.
This phase matching is realized by the passage of quantized vortices across the local superflow.
Phase slippages produce a change in the superfluid kinetic energy\cite{AndersonRMP1966}, 
which should be detected in the Helmholtz resonance\cite{AvenelPRL1985}.
We will discuss the dissipation in terms of the phase matching process\cite{Tani2021}.
  
In summary, we have precisely determined the superfluid critical exponent $\zeta$ for ${}^{\rm 4}$He confined in porous Gelsil glasses with pore sizes of 3.0 nm using the Helmholtz resonator technique. 
$\zeta$ is found to be $1.0 \pm 0.1$ for all the pressures realized in this technique, $0.1 < P < 2.4$ MPa. 
The value of the exponent provides decisive evidence for the 4D XY quantum criticality realized not only at 0 K QCP, but at finite temperatures.
To our knowledge, $^4$He in Gelsil provides the first example of a 4D XY bosonic system in real matter.
Further studies of QPT in $^4$He in other nanoporous media will stimulate theoretical studies of 4D XY in the Bose-Hubbard model\cite{ProsniakSciRep2019} and 4D XY QPTs proposed in fermionic systems\cite{4DXYHTSC}.

\end{document}